\documentclass{article}
\newcommand{\comment}[1]{}
\newcommand{\ignore}[1]{}
\ignore{\makeatletter
\let\proof\@undefined
\let\endproof\@undefined
\let\theorem\@undefined
\let\definition\@undefined
\let\endtheorem\@undefined
\let\enddefinition\@undefined
\let\remark\@undefined
\let\endremark\@undefined
\let\c@theorem\@undefined
\let\claim\@undefined
\let\proposition\@undefined
\let\lemma\@undefined
\let\corollary\@undefined
\makeatother}

\usepackage{amssymb}
\ignore{\setcounter{tocdepth}{3}}
\usepackage{graphicx}
\usepackage{fullpage}
\usepackage{url}

\ignore{}

\usepackage{amsmath,amssymb,amsthm,algorithm,algorithmic,prettyref,boxedminipage}
\usepackage{pstricks,pst-node,url}
\usepackage{xcolor}
\definecolor{foocite}{rgb}{0,0.75,0}
\definecolor{foolink}{rgb}{0,0,1}
\definecolor{foourl}{rgb}{1,0,0}
\usepackage[colorlinks=true,citecolor=foocite,linkcolor=foolink,urlcolor=foourl]{hyperref}
\newtheorem{theorem}{Theorem}
\newrefformat{theorem}{Theorem~\ref{#1}}
\newrefformat{eq}{Equation~\eqref{#1}}
\newrefformat{chap}{Chapter~\ref{#1}}
\newrefformat{fig}{Figure~\ref{#1}}
\def\xthm[#1][#2][#3]{\newtheorem{#2}[theorem]{#3} \newrefformat{#2}{#3 \ref{#11}}}
\xthm[#][defn][Definition]
\xthm[#][claim][Claim]
\xthm[#][conj][Conjecture]
\xthm[#][prop][Proposition]
\xthm[#][lemma][Lemma]
\xthm[#][eg][Example]
\xthm[#][fact][Fact]
\xthm[#][cor][Corollary]
\xthm[#][app][Appendix]
\xthm[#][alg][Algorithm]
\xthm[#][step][Step]
\xthm[#][remark][Remark]

\newcommand{\OPT}{\ensuremath{\mathrm{OPT}}}
\newcommand{\R}{\ensuremath{\mathbb{R}}}

\newcommand{\NP}{\ensuremath{\mathsf{NP}}}
\newcommand{\PP}{\ensuremath{\mathsf{P}}}
\newcommand{\APX}{\ensuremath{\mathsf{APX}}}

\newcommand{\bs}{\backslash}

\begin{document}


\title{$k$-Edge-Connectivity: Approximation and LP Relaxation}
\author{David Pritchard\thanks{'Ecole Polytechnique F\'ed\'erale de Lausanne; partially supported by an NSERC post-doctoral fellowship}}
\date{\today}

\maketitle
\begin{abstract}In the $k$-edge-connected spanning subgraph problem we are given a graph $(V, E)$ and costs for each edge, and want to find a minimum-cost $F \subset E$ such that $(V, F)$ is $k$-edge-connected. We show there is a constant $\epsilon>0$ so that for all $k>1$, finding a $(1+\epsilon)$-approximation for $k$-ECSS is $\NP$-hard, establishing a gap between the unit-cost and general-cost versions.
Next, we consider the \emph{multi-subgraph} cousin of $k$-ECSS, in which we purchase a \emph{multi-subset} $F$ of $E$, with unlimited parallel copies available at the same cost as the original edge.
We conjecture that a $(1 + \Theta(1/k))$-approximation algorithm exists, and we describe an approach based on graph decompositions applied to its natural linear programming (LP) relaxation. The LP is essentially equivalent to the Held-Karp LP for TSP and the undirected LP for Steiner tree. We give a family of extreme points for the LP which are more complex than those previously known.
\end{abstract}

\newcommand{\LPNKb}[1]{\ensuremath{\mathrm{(}\mathcal{N}'_{#1}\mathrm{)}}}
\newcommand{\LPNKu}[1]{\ensuremath{\mathrm{(}\mathcal{N}_{#1}\mathrm{)}}}


\section{Introduction}
In the $k$-edge-connected spanning subgraph problem ($k$-ECSS), we are given an input graph $G$ with edge costs, and must select a minimum-cost subset of edges so that the resulting graph has edge-connectivity $k$ between all vertices. This is a natural problem for applications, since it is the same as seeking resilience against $(k-1)$ edge failures, or the ability to route $k$ units of flow between any pair of vertices. A natural variant of $k$-ECSS is to allow each edge to be purchased repeatedly, as many times as desired, with each copy at the same cost. We call this the \emph{$k$-edge-connected spanning multi-subgraph problem} ($k$-ECSM).

When $k=1$ the $k$-ECSS and $k$-ECSM problems are both equivalent to the \emph{minimum spanning tree} problem, which is well-known to be solvable in polynomial time, but they are non-trivial for $k>1$. We consider \emph{approximation algorithms} for these problems: an algorithm that approximately solves $k$-ECSS or $k$-ECSM is said to be an \emph{$\alpha$-approximation}, or have \emph{approximation ratio $\alpha$}, if it always outputs a solution with cost at most $\alpha$ times optimal.

Here we survey the oldest and newest results for $k$-ECSM and $k$-ECSS. Frederickson \& J\'{a}j\'{a} gave a $2$-approximation algorithm for 2-ECSS~\cite{FJ81}, and a $3/2$-approximation in the special case of metric costs~\cite{FJ82}. A $3/2$-approximation is possible for 2-ECSM~\cite{BC08}. For $k$-ECSS/$k$-ECSM in general, Khuller \& Vishkin~\cite{KV94} gave a matroid-based 2-approximation, and Jain's iterated LP rounding framework~\cite{J98} also gives a 2-approximation. Goemans \& Bertsimas~\cite{GB93} give an approximation algorithm for $k$-ECSM with ratio $\frac{3}{2}$ when $k$ is even, and $(\frac{3}{2}+\frac{1}{2k})$ when $k$ is odd. Fernandes~\cite{Fern98} showed 2-ECSS is \APX-hard, even for unit costs.

An important special case is where all edges have unit cost. Then $k$-ECSS gets \emph{easier} to approximate as $k$ gets larger: Gabow et al.~\cite{GGTW09} gave an elegant $(1+2/k)$-approximation algorithm for $k$-ECSS/$k$-ECSM using iterated LP rounding, and they showed that for some fixed $\epsilon>0$, for all $k>1$, it is $\NP$-hard to get a $(1+\epsilon/k)$-approximation algorithm for unit-cost $k$-ECSS. Together, these establish a $1+\Theta(1/k)$ approximability threshold for unit-cost $k$-ECSS. Improvements to the constant, and improvements in the special case that the input graph is simple, appear in Cheriyan \& Thurimella~\cite{CT00} and Gabow \& Gallagher~\cite{GG08}.


\subsection{Contributions}
\subsubsection{Hardness Results (\prettyref{sec:kecss-hard})}
Our first main result is the following hardness for $k$-ECSS:
\begin{theorem}\label{thm:kecss-hard}
There is a constant $\epsilon>0$ so that for all $k \ge 2$, it is \NP-hard to approximate $k$-ECSS within ratio $1+\epsilon$, even if the costs are $0$-$1$.
\end{theorem}
\newtheorem*{theorem-main1}{\prettyref{thm:kecss-hard}}
Although $\epsilon \approx \frac{1}{300}$ here is small, the qualitative difference is important: whereas the approximability of unit-cost $k$-ECSS tends to 1 as $k$ tends to infinity, we see that the approximability of general-cost $k$-ECSS is bounded away from 1.

Next we establish a relatively straightforward hardness result for $k$-ECSM.
\begin{prop}\label{prop:2ecsm-hard}
The 2-ECSM problem is $\APX$-hard.
\end{prop}
The key step is to show that 2-ECSM and \emph{metric 2-ECSS} are basically the same problem.
First, we use the following well-known fact: in $k$-ECSM, the input is metric without loss of generality~\cite{GB93} (i.e.\ the graph is complete and its costs satisfy the triangle inequality).\footnote{To see this, take the metric closure (i.e.\ shortest path costs), solve it, and replace each $uv$-edge in the solution with a shortest $u$-$v$ path from the original graph; it is not hard to show this preserves $k$-edge-connectivity. In $k$-ECSS, note metricity is not WOLOG, since the replacement step here can introduce multiple edges.} Then, simple reduction techniques show that under metric costs, any 2-ECSM can be efficiently converted to a 2-ECSS without increasing the costs. We remark that this approach also yields a simpler $3/2$-approximation for 2-ECSM (c.f.\ \cite{BC08}), using the $3/2$-approximation for metric 2-ECSS~\cite{FJ82} as a black box.

What \prettyref{prop:2ecsm-hard} leaves to be desired is hardness for $k$-ECSM, $k>2$, and asymptotic dependence on $k$. Why is it hard to show these problems are hard? The hard instances for $k$-ECSS given by \prettyref{thm:kecss-hard} and \cite{GGTW09} contain certain \emph{mandatory parts} that are ``without loss of generality" included in the optimal feasible solution; the argument proceeds to show hardness of the residual problem once the mandatory parts are included. But coming up with suitable mandatory parts for $k$-ECSM, while keeping the residual problem hard, is tricky: e.g.\ the proof of \prettyref{thm:kecss-hard} will use a spanning tree of zero-cost edges, but in $k$-ECSM this leads to a trivial instance (buy that spanning tree $k$ times). The known hardness for $k$-VCSS (vertex connectivity) by Kortsarz et al.\ \cite{KKL04} is similar: we take hard instances of 2-VCSS and add $(k-2)$ new vertices, connected to all other vertices by 0-cost (mandatory) edges. A new trick seems to be needed to get a good hardness result for $k$-ECSM.

\subsubsection{$k$-ECSM Conjecture (\prettyref{sec:conjer})}
We conjecture that approximation ratio $1+O(1/k)$ should be possible for $k$-ECSM, using LPs. Obtain the natural LP relaxation of $k$-ECSM by allowing edges to be purchased fractionally: introduce a variable $x_e$ for each edge, and require that there is a fractional value of at least $k$ spanning each cut (see Figure~\ref{fig:kecss}, where $\delta(S)$ denotes the set of edges with exactly one end in $S$).
\begin{figure}
\hfill\begin{minipage}{5cm}
\begin{align}
\min \Big\{\sum_{e \in E} c_ex_e: \quad& x \in \R^E \tag{\ensuremath{\mathcal{N}_k}}\label{eq:LP-NKu}
\\
\sum_{e \in \delta(S)} x_e \ge k, \quad&\forall \varnothing \neq S \subsetneq V 
 \notag \\
x_e \ge 0, \quad& \forall e\in E 
\Big\} \notag
\end{align}
\end{minipage}
\hfill \vline \hfill
\begin{minipage}{5cm}
\begin{align}
\min \Big\{\sum_{e \in E} c_ex_e: \quad& x \in \R^E \tag{\ensuremath{\mathcal{N}'_k}}\label{eq:LP-NKb} \\
\sum_{e \in \delta(v)} x_e = k, \quad&\forall v \in V 
\notag\\
\sum_{e \in \delta(S)} x_e \ge k, \quad&\forall \varnothing \neq S \subsetneq V 
\notag \\
x_e \ge 0, \quad& \forall e\in E \label{eq:LP-NKb3} \Big\} \notag
\end{align}
\end{minipage}
\hfill\caption{
The undirected relaxation for $k$-edge connected spanning multi-subgraph. The unbounded version \eqref{eq:LP-NKu} is on the left, the bounded version \eqref{eq:LP-NKb} is on the right. They have the same value for metric costs, including all $k$-ECSM instances.}\label{fig:kecss}
\end{figure}

\begin{conj}\label{conj:kecss}
There is a polynomial-time approximation algorithm for $k$-ECSM which produces a solution of value at most $(1+C/k)\cdot \OPT\eqref{eq:LP-NKu}$ for some universal constant $C$.
\end{conj}
\newtheorem*{conj:main2}{\prettyref{conj:kecss}}
This conjecture implies a $(1+C/k)$-approximation algorithm, since $\OPT\eqref{eq:LP-NKu}$ is a lower bound on the optimal $k$-ECSM cost.
What makes us think \prettyref{conj:kecss} is true? First, we know it holds for unit costs. Second, the same holds in related high-\emph{width} problems; to explain, say an integer program has width $W$ if in every constraint, the right-hand side is at least $W$ times every coefficient. Multicommodity flow/covering problems in trees are closely related to \eqref{eq:LP-NKu} via \emph{uncrossing}~(e.g.\ \cite{J98,GGTW09,GG08}) and they admit an LP-based $1+O(1/W)$-approximation algorithm~\cite{KPP1x} (in that setting $W$ is the minimum edge capacity). Similar phenomena are known for LP relaxations of other structured integer programs~\cite{CMS07,CEK09,PC10,BKN09}. In $k$-ECSM the width is $k$ so one may view our conjecture as seeking integrality gap\footnote{The integrality gap is the worst-case ratio of the integral optimum to the LP optimum.} and approximation ratio $1+O(1/W)$.

Later, we show an open problem of~\cite{BJY04} --- can every $k$-edge connected graph be partitioned into two spanning $(\frac{k}{2}-O(1))$-edge-connected subgraphs? --- would imply a nonconstructive version of \prettyref{conj:kecss}. Few partial results towards \prettyref{conj:kecss} are known: the integrality gap of \LPNKu{1} is $2(1-1/n)$~\cite{GB93}, and that of \LPNKu{2} is at most 3/2~\cite{Wo80}. For general $k$, the best integrality gap bounds known for \eqref{eq:LP-NKu} come from the approximation algorithms \cite{J98,GB93,GG08,GGTW09} mentioned earlier.

One further motivation to investigate the conjecture has to do with the \emph{parsimonious property} of Goemans \& Bertsimas~\cite{GB93}. Using metricity and \emph{splitting-off}, they showed the constraint $\forall v\in V: x(\delta(v))=k$ can be added to \eqref{eq:LP-NKu} without affecting the value of the LP (the strengthened LP \eqref{eq:LP-NKb} is shown in Figure~\ref{fig:kecss}). As observed in~\cite{GB93}, parsimony implies that \prettyref{conj:kecss} would give a $(1+\frac{C}{k})$-approximation algorithm for \emph{subset $k$-ECSM}, where we require edge-connectivity $k$ only amongst a pre-specified set of terminal nodes (generalizing the Steiner tree problem). Thus even if we don't care about LPs \emph{a priori}, they have algorithmic dividends in \prettyref{conj:kecss}.

\comment{\begin{table}[ht]
\begin{center}
\begin{tabular}{||c|cc|ccc||}
\hline \hline
 & \multicolumn{2}{c}{Unit Costs} & \multicolumn{2}{c}{Arbitrary Costs}&\\
 & lower bound & upper bound & lower bound & upper bound&\\
\hline
$k$-ECSM   & $1+\epsilon$ for $k=2$ & $1+O(1/k)$ & $1+\epsilon$ for $k=2$ & $\frac{3}{2}\{+\frac{1}{2k}\}_{\textrm{odd $k$}}$ \& Conj.\ \ref{conj:kecss}& \\
$k$-ECSS & $1+\epsilon/k$ & $1+O(1/k)$ & $1+\epsilon$~(Thm.\ \ref{thm:kecss-hard}) & 2& \\
\hline \hline
\end{tabular} \caption{Approximability of $k$-ECSS and $k$-ECSM. Note that all the upper bounds can be obtained by \eqref{eq:LP-NKu}-relative approximation algorithms. Each $\epsilon$ represents a small fixed constant independent of $k$.}\label{tab:kecss}
\end{center}
\end{table}}

\subsubsection{Complex Extreme Points (\prettyref{sec:hk-cons})}\label{sec:hk-ext}
In both of the LPs \eqref{eq:LP-NKu} and \eqref{eq:LP-NKb}, note that $k$ serves only as a scaling factor: $x$ is feasible for \LPNKu{1} iff $kx$ is feasible for $\eqref{eq:LP-NKu}$. In fact, these LPs are well-studied: \LPNKu{1} is equivalent (by the parsimonious property~\cite{GB93}) to the \emph{undirected cut relaxation of the Steiner tree problem} and \LPNKb{2} is the \emph{Held-Karp relaxation of the Traveling Salesman Problem}. We demonstrate a family of extreme point solutions to these ubiquitous LPs which are more complex than were previously known.

For a solution $x$, the \emph{support} is the edge set $\{e \mid x_e > 0\}$, and the \emph{support graph} is the graph with vertex set $V$ and the support for its edge set.  The \emph{fractionality} of $x$ is $\min \{x_e \mid e \in E, x_e > 0\}$.
\begin{theorem}\label{theorem:family}
There are extreme point solutions for the linear program \LPNKb{2} with fractionality exponentially small in $|V|$, and whose support graph has maximum degree linear in $|V|$.
\end{theorem}
The members of the family are also extreme point solutions for \LPNKu{2}, since \LPNKb{2} is a face of \LPNKu{2}.
The motivation for this theorem comes from a common design methodology in LP-based approximation algorithms~\cite{J98,GGTW09,Go06,SL07}: algorithmically exploit good properties of extreme point solutions. E.g., Jain's algorithm~\cite{J98} uses the fact that when \eqref{eq:LP-NKu} is generalized to skew-submodular connectivity requirements, every extreme solution $x^*$ has an edge $e$ with $x^*_e \ge \frac{1}{2}$. Hence, complex extreme points give some idea of what properties might or might not exist that can be exploited algorithmically.

\prettyref{theorem:family} significantly improves previous results in the same vein. (A long-standing conjecture that the Held-Karp relaxation \LPNKb{2} has integrality gap at most $4/3$ has motivated some of the work, e.g.\ \cite{CV00a,BB08}.) Boyd and Pulleyblank~\cite{BP91} showed that for any even $|V| \ge 10$, there is an extreme point of \LPNKb{2} with values in $\{1/t, 2/t, 1-2/t, 1-1/t, 1\}$, hence fractionality $2/(|V|-4)$.
Cheung~\cite{Cheung03} gave a family of extreme points on $\Theta(t^2)$ vertices with maximum degree $4t+2$ and entries in $\{1/(2t+1), 1-1/(2t+1), 1\}$, for every integer $t \ge 1$, hence maximum degree $\Theta(\sqrt{|V|})$ in the support graph.
The construction in \prettyref{theorem:family} was found with the assistance of computational methods, as we describe later.

\ignore{can have arbitrarily high maximum degree. He gave a family of extreme points on $\Theta(t^2)$ vertices with maximum degree $4t+2$ and entries in $\{1/(2t+1), 1-1/(2t+1), 1\}$, for every integer $t \ge 1$, hence maximum degree $\Theta(\sqrt{|V|})$ in the support graph.}

We remark that a manuscript of Cunningham \& Zhang~\cite{CZ92} observes that by gluing together copies of the Boyd-Pulleyblank construction as blocks (2-vertex-connected components), one can get extreme points of \LPNKu{2} with \emph{denominator} $\min\{t \in \mathbb{Z}_{\ge 0} \mid tx \textrm{ integral}\}$ of value $\Omega(\sqrt{|V|}!)$. However, the fractionality is no worse than that of the Boyd-Pulleyblank construction, and gluing does not work for \LPNKb{2}. Plus, insofar as we care about designing approximation algorithms, we may well solve $k$-ECSS separately on each block, so this does not shed light on limits of the LP-based approach.



\section{Hardness Results}\label{sec:kecss-hard}
In our hardness theorem for $k$-ECSS, we reduce from the following problem. (Here $\uplus$ denotes disjoint union.)
\begin{center}\begin{boxedminipage}{12cm}
\noindent \textsc{Path-Cover-of-Tree}
\\ Input: A tree $T=(V, E)$ and another set $X \subset \tbinom{V}{2}$ of edges/pairs.
\\ Output: A subset of $Y$ of $X$ so that $(V, E \uplus Y)$ is 2-edge-connected.
\\ Objective: Minimize $|Y|$.
\end{boxedminipage}
\end{center}
\textsc{Path-Cover-of-Tree} is sometimes called the \emph{tree augmentation problem} and a 1.8-approximation is published~\cite{GFKZ09}; as an aside, it is basically equivalent to the special case of 2-ECSS where the input graph contains a connected subgraph of cost zero, plus some unit-cost edges. We give it the alternate name \textsc{Path-Cover-of-Tree} because it is more natural for us to interpret it as covering a tree's edges with a minimum-size subcollection of a given collection of paths. To make this explicit, for an edge $x = \{u, v\} \in X$ let $P_x$ denote the edges of the unique $u$-$v$ path in $T$. We rehash the proof of the following proposition since we will recycle its methodology.
\begin{prop}[folklore]\label{prop:folklore}
$Y$ is feasible for \textsc{Path-Cover-of-Tree} if and only if $\bigcup_{x \in Y} P_x = E$.
\end{prop}
\begin{proof}
For every edge $e$ of $T$, a \emph{fundamental cut} of $e$ and $T$ means the vertex set of either connected component of $T \bs e$.

Let $\delta_F(U)$ denote $\delta(U)$ in the graph $(V, F)$. First, $Y$ is feasible if $|\delta_{E \uplus Y}(U)| \ge 2$ for every set $U$ with $\varnothing \neq U \subsetneq V$. But $|\delta_E(U)|$ is 1 when $U$ is a fundamental cut and at least 2 otherwise; hence $Y$ is feasible iff $|\delta_Y(U)| \ge 1$ for every fundamental cut $U$.

Second, when $U$ is a fundamental cut, say for an edge $e \in E$, $|\delta_Y(U)| \ge 1$ iff $\bigcup_{x \in Y} P_x$ contains $e$. Taking this together with the previous paragraph, we are done.
\end{proof}
\textsc{Path-Cover-of-Tree} is shown \NP-hard in \cite{FJ81} and a similar construction implies \APX-hardness --- we give the proof in the appendix. As an aside, it is even hard for trees of depth 2; compare this with the depth-1 instances which are in \PP\ since they can be shown isomorphic to \emph{edge cover}. Now we prove the main hardness result:
\begin{theorem-main1}
Let it be $\NP$-hard to approximate \textsc{Path-Cover-of-Tree} within ratio $1+\epsilon$. Then for all integers $k \ge 2$, it is \NP-hard to approximate $k$-ECSS within ratio $1+\epsilon$, even for $0$-$1$ costs.
\end{theorem-main1}
\begin{proof}
Let $(T=(V, E), X)$ denote an instance of \textsc{Path-Cover-of-Tree}. We construct a $k$-ECSS instance on the same vertex set, with edge set $F$. For each $e \in E$, we put $k-1$ zero-cost copies of the edge $e$ into $F$. For each $x \in X$, put one unit-cost copy of the edge $x$ into $F$. These are all the edges of $F$; and although $(V, F)$ is a multigraph, we later show that this can be avoided.

First we show the multigraph instance is hard. Clearly, there is an optimal solution for the $k$-ECSS instance which includes all copies of the 0-cost edges. Let $(k-1)E$ denote these 0-cost edges. The same logic as in the proof of \prettyref{prop:folklore} (analysis using fundamental cuts) shows that $Y$ is a feasible solution for the \textsc{Path-Cover-of-Tree} instance if and only if $(k-1)E \uplus Y$ is a feasible solution for the $k$-ECSS instance. Since costs are preserved between the two problems, it follows that an $\alpha$-approximation algorithm for $k$-ECSS would also give an $\alpha$-approximation algorithm for \textsc{Path-Cover-of-Tree}, and we are done.

Finally, here is how we make $(V, F)$ a simple graph: replace every vertex $v \in V$ of the tree by a $(k+1)$-clique of 0-cost edges; replace every edge $uv \in E$ of the tree by any $k-1$ zero-cost edges between the two cliques for $u$ and $v$; replace each edge $x \in X$ by any unit-cost edge between the cliques for $u$ and $v$. We proceed similarly to before: when $U$ is a vertex set of the newly constructed graph, we see $\delta(U)$ has at least $k$ 0-cost edges unless $U$ is a ``blown-up" version of a fundamental cut (i.e., unless there is a fundamental cut $U_0$ of $T$ so that $U$ exactly equals the set of vertices in cliques corresponding to $U_0$). As before, the residual problem assuming these edges are bought is the same as the instance $(T, X)$ (in a cost-preserving way), so we are done.
\end{proof}

\subsection{Hardness of 2-ECSM (Proof of \prettyref{prop:2ecsm-hard})}
To show that 2-ECSM is \APX-hard, we prove that it is ``the same" as metric 2-ECSS, i.e.\ the special case of 2-ECSS on complete metric graphs. Metric 2-ECSS is \APX-hard by a general result of \cite{BBH+04}\footnote{
Here is a sketch for the reader, somewhat simpler than the more general results of \cite{BBH+04}. Take a family of hard TSP instances with costs 1 and 2~\cite{PY93}. Using a little case analysis, \cite{BBH+04} shows that a 2-ECSS can be transformed to a Hamiltonian cycle (TSP tour) by repeatedly replacing two edges with one edge, which does not incerase the overall cost if edge costs are 1 and 2; so for these instances, TSP and 2-ECSS are the same. In particular on these (metric) instances, finding the min-cost 2-ECSS is \APX-hard.} and so this gives us what we want. The key observation is the following.

\begin{prop}\label{prop:convert}
In a metric instance, given a 2-ECSM $(V, F)$, we can obtain in polynomial time a 2-ECSS $(V, F')$ with $c(F') \le c(F)$, as long as $|V| \ge 3$.
\end{prop}
\noindent In other words, parallel edges can be eliminated without increasing the cost. (A similar observation in \cite{FJ82} turns a 2-ECSS into a 2-VCSS for metric instances.) 
\begin{proof}
We may assume $F$ is minimal, i.e.\ that deleting any edge from $(V, F)$ leaves a non-2-edge-connected graph. This implies there are no parallel triples. Next, suppose there is a parallel pair between some vertices $u$ and $v$. If there is any $u$-$v$ path not using a $uv$ edge, it is easy to see that deleting one of the parallel $uv$ edges contradicts minimality. Therefore we may assume $uv$ is a cut edge (bridge) of the simplification of $(V, F)$; call this the \emph{bridge assumption}.

Since the graph is connected and $|V| \ge 3$, at least one of $u$ or $v$ (say $u$ WOLOG) has another neighbour $w$. By the bridge assumption $v$ is not adjacent to $w$. We will argue that the set $F'$ obtained by deleting a $uv$ edge, a $uw$ edge, and adding a $vw$ edge, is still 2-edge-connected. Iterating this operation we are done (since the cost does not increase and the number of parallel pairs decreases each time).

Since $(V, F)$ is 2-edge-connected, it has a $u$-$w$ path $P$ not using the deleted $uw$ edge. By the bridge assumption, $P$ does not use any $uv$ edge. Note that $|\delta_{F'}(S)| < |\delta_F(S)|$ only if $S$ contains $v$ and $w$ but not $u$ (or vice-versa). But then $\delta_{F'}(S)$ contains the remaining $uv$ edge and at least one edge from $P$. So $|\delta_{F'}(S)| \ge 2$ for all $\varnothing \neq S \subsetneq V$ and we are done.
\end{proof}


\begin{proof}[Proof of \prettyref{prop:2ecsm-hard}]
Since metric 2-ECSS is \APX-hard~\cite{BBH+04}, it is enough to show that any $\alpha$-approximation algorithm for 2-ECSM gives an $\alpha$-approximation for metric 2-ECSS. The metric 2-ECSS algorithm is: compute an $\alpha$-approximately-optimal 2-ECSM $F$ and apply \prettyref{prop:convert} to get a 2-ECSS $F'$ with $c(F') \le c(F)$. Using \prettyref{prop:convert} a second time, and using the fact that every 2-ECSS is trivially a 2-ECSM, we see the optimal 2-ECSS and 2-ECSM values are the same. Hence $F'$ is an $\alpha$-approximately-optimal 2-ECSS, as needed.
\end{proof}

\section{$k$-ECSM Conjecture and Connectivity Decomposition}\label{sec:conjer}
Here is the conjecture made in the introduction. We will relate it to questions about graph decomposition.
\begin{conj:main2}
There is a polynomial-time approximation algorithm for $k$-ECSM which produces a solution of value at most $(1+C/k)\cdot \OPT\eqref{eq:LP-NKu}$ for some universal constant $C$.
\end{conj:main2}

For positive integers $A$ and $B$, define $f(A, B)$ to be the least integer $f$ so that every $f$-edge-connected multigraph can be partitioned into two spanning subgraphs, one $A$-edge-connected and one $B$-edge-connected.
Bang-Jensen and Yeo~\cite{BJY04} ask the following question, which we call the \emph{splitting hypothesis}: is there a constant $C$ such that $f(k, k) \le 2k+C$ for all integers $k$? It has consequences for \prettyref{conj:kecss}:
\begin{theorem}\label{theorem:interesting}
If the splitting hypothesis holds, then every $k$-ECSM instance has a solution with cost at most $(1+C/k)\cdot\OPT\eqref{eq:LP-NKu}$, i.e.\ the integrality gap of \eqref{eq:LP-NKu} is at most $1+C/k$.
\end{theorem}
\noindent This would not prove \prettyref{conj:kecss} due the lack of a polynomial-time algorithm; but one might guess that once the core combinatorial problem is solved, a polynomial-time implementation could be found, as happened in~\cite{CMS07}.

Before proving \prettyref{theorem:interesting} we make some other remarks about $f$. The Nash-Williams/Tutte theorem implies $f(A, B) \le 2(A+B)$. The lower bound $f(A, B) \ge A+B$ is very easy, by considering an $(A+B-1)$-regular, $(A+B-1)$-edge-connected graph. This lower bound can be raised by 1 or 2 in a few cases, e.g.~the fact that a spanning tree has average degree almost 2 implies that $f(A, 1) \ge A+2$ and $f(1, 1) \ge 4$. An example of another small improvement is that any $d$-regular, $d$-edge-connected graph with no Hamilton path implies $f(d-2, 1) \ge d+1$ since, were it to contain a spanning tree disjoint from a $(d-2)$-edge-connected subgraph, that tree would have maximum degree 2 and hence be a Hamilton path. Such a graph is known to exist at least when $d=3$~\cite[Fig.~5.4]{Si05} and by taking parallel copies we also get one for any $d$ which is a multiple of 3. For any $d \ge 3$ there is a $d$-regular, $d$-edge-connected graph with no Hamiltonian cycle~\cite{Mer73}, similarly implying $f(d-2, 2) \ge d+1$.

It seems the only value of $f$ known exactly is $f(1, 1)=4$. M. DeVos asked\footnote{\url{http://garden.irmacs.sfu.ca/?q=op/partitioning_edge_connectivity}} online whether $\forall A, B : f(A, B) \le A+B+2$ holds, which is still open.

\begin{proof}[Proof of \prettyref{theorem:interesting}]
Let $x^*$ be an optimal extreme point solution to \eqref{eq:LP-NKu}. Since $x^*$ is rational, there is an integer $t$ such that $tx^*$ is integral. Then, it is easy to see that $tx^*$ (or more precisely, the multigraph obtained by taking $tx^*_e$ copies of each edge $e$) is a $tk$-edge-connected spanning multisubgraph. Likewise, for any positive integer $\alpha$, $\alpha tx^*$ is a $(\alpha tk)$-ECSM.

By induction, the splitting hypothesis easily gives the following.
\begin{claim}
For all positive integers $k$ and $n$, every $(2^n(k+C) - C)$-ECSM can be decomposed into $2^n$ disjoint $k$-ECSMs.
\end{claim}
Now, for any integer $n$, let us pick $\alpha$ just large enough that $\alpha tk \ge (2^n(k+C) - C)$. Therefore, $\alpha tx^*$ can be decomposed into $2^n$ disjoint $k$-ECSMs. The cheapest one has cost at most
$$\frac{c(\alpha tx^*)}{2^n} = \alpha t 2^{-n} c(x^*) = \Bigl\lceil \frac{2^n(k+C) - C}{tk} \Bigr\rceil t 2^{-n} \OPT\eqref{eq:LP-NKu}.$$
Then using $\lceil \frac{2^n(k+C) - C}{tk} \rceil \le \lceil \frac{2^n(k+C)}{tk} \rceil \le \frac{2^n(k+C)}{tk}+1$, we see there is a $k$-ECSM with cost at most
$$\Bigl(\frac{2^n(k+C)}{tk}+1\Bigr) t 2^{-n} \OPT\eqref{eq:LP-NKu} = (1 + C/k + t/2^n)\OPT\eqref{eq:LP-NKu}.$$
This establishes that the integrality gap is no more than $1 + C/k + t/2^n$. Taking $n \to \infty$, we are done (since the integrality gap is some fixed real, and since $t$ doesn't depend on $n$).
\end{proof}

We feel strongly that the following holds.
\begin{conj}
$f(A, 1) = A + o(A).$
\end{conj}
\noindent For example, given a 100-edge-connected graph, if we want to delete a spanning tree of our choice and keep high edge-connectivity, 49 hardly seems like the best possible. It is not too hard to see (using repeated splitting and merging) that the splitting hypothesis would imply $f(A, 1) = A + O(C \ln A)$ and hence prove this conjecture.

Variants of $f$ have received some attention. For edge-connectivity in hypergraphs, $f(1, 1)$ is not finite~\cite{BT03}. It is not known whether the analogue of $f(1, 1)$ in \emph{directed graphs} is finite~\cite{BJY04,BJ09}.

\section{Complex Extreme Points for \LPNKb{2}}\label{sec:hk-cons}
Now we give our construction of a new family of extreme points for the TSP subtour relaxation \LPNKb{2}; as mentioned earlier, it can be scaled by $k/2$ to give an extreme point for \eqref{eq:LP-NKb} or \eqref{eq:LP-NKu}, which is relevant to LP-based approaches for $k$-ECSM.

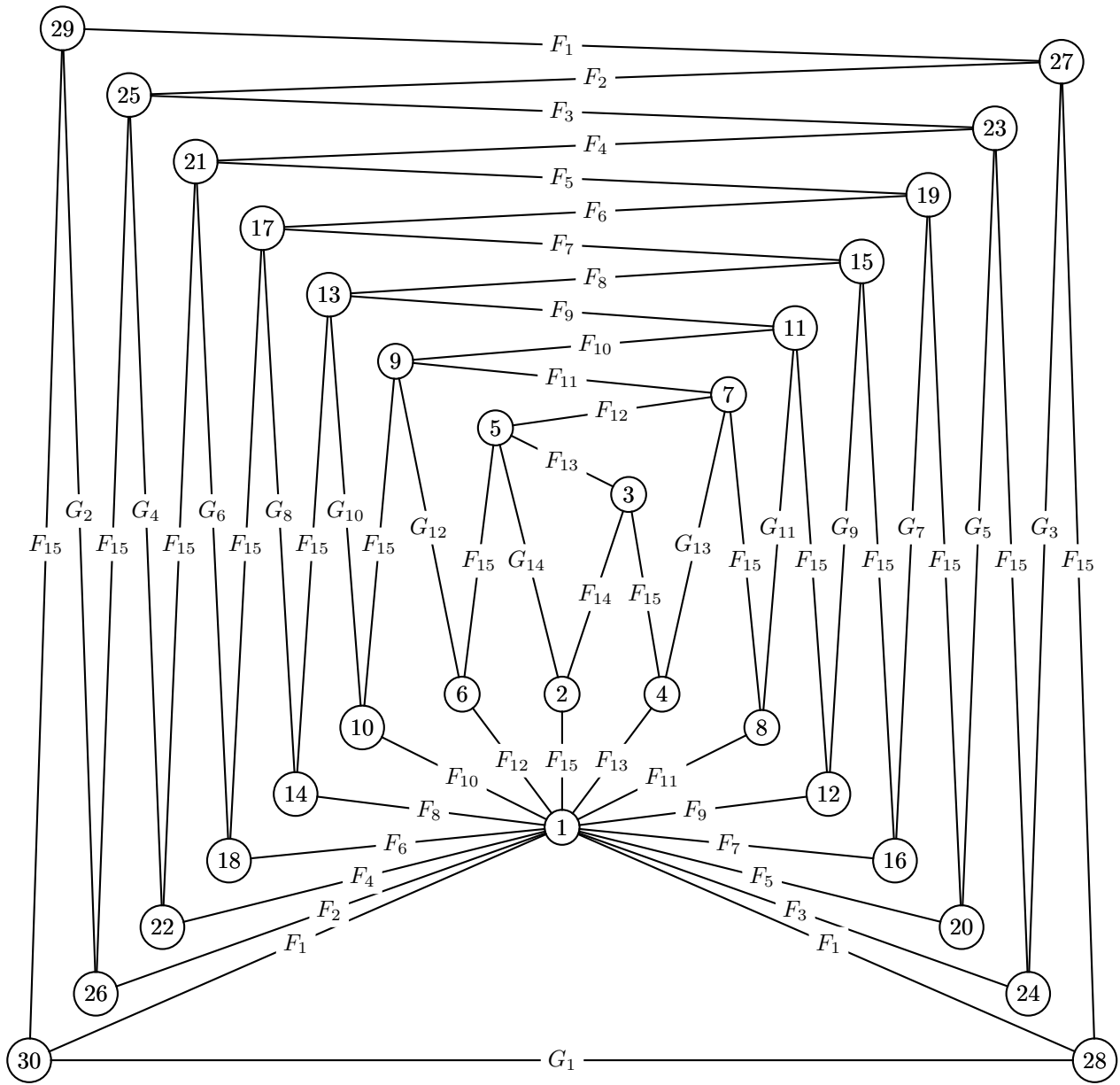
\begin{figure}
\begin{center}
\begin{pspicture}(-4.5,-12)(11.5,5.5)
\psset{unit=0.25cm}
\cnode(-18,-42){3pt}{P30A}
\cnode(-16,20){3pt}{P30B}
\ncline{P30A}{P30B}
\mput*{$F_{15}$}
\cnode(-18,-42){3pt}{P31A}
\cnode(14,-28){3pt}{P31B}
\ncline{P31A}{P31B}
\mput*{$F_{1}$}
\cnode(-14,-38){3pt}{P32A}
\cnode(-16,20){3pt}{P32B}
\ncline{P32A}{P32B}
\mput*{$G_{2}$}
\cnode(-14,-38){3pt}{P33A}
\cnode(-12,16){3pt}{P33B}
\ncline{P33A}{P33B}
\mput*{$F_{15}$}
\cnode(-14,-38){3pt}{P34A}
\cnode(14,-28){3pt}{P34B}
\ncline{P34A}{P34B}
\mput*{$F_{2}$}
\cnode(-10,-34){3pt}{P35A}
\cnode(-12,16){3pt}{P35B}
\ncline{P35A}{P35B}
\mput*{$G_{4}$}
\cnode(-10,-34){3pt}{P36A}
\cnode(-8,12){3pt}{P36B}
\ncline{P36A}{P36B}
\mput*{$F_{15}$}
\cnode(-10,-34){3pt}{P37A}
\cnode(14,-28){3pt}{P37B}
\ncline{P37A}{P37B}
\mput*{$F_{4}$}
\cnode(-6,-30){3pt}{P38A}
\cnode(-8,12){3pt}{P38B}
\ncline{P38A}{P38B}
\mput*{$G_{6}$}
\cnode(-6,-30){3pt}{P39A}
\cnode(-4,8){3pt}{P39B}
\ncline{P39A}{P39B}
\mput*{$F_{15}$}
\cnode(-6,-30){3pt}{P40A}
\cnode(14,-28){3pt}{P40B}
\ncline{P40A}{P40B}
\mput*{$F_{6}$}
\cnode(-2,-26){3pt}{P41A}
\cnode(-4,8){3pt}{P41B}
\ncline{P41A}{P41B}
\mput*{$G_{8}$}
\cnode(-2,-26){3pt}{P42A}
\cnode(0,4){3pt}{P42B}
\ncline{P42A}{P42B}
\mput*{$F_{15}$}
\cnode(-2,-26){3pt}{P43A}
\cnode(14,-28){3pt}{P43B}
\ncline{P43A}{P43B}
\mput*{$F_{8}$}
\cnode(2,-22){3pt}{P44A}
\cnode(0,4){3pt}{P44B}
\ncline{P44A}{P44B}
\mput*{$G_{10}$}
\cnode(2,-22){3pt}{P45A}
\cnode(4,0){3pt}{P45B}
\ncline{P45A}{P45B}
\mput*{$F_{15}$}
\cnode(2,-22){3pt}{P46A}
\cnode(14,-28){3pt}{P46B}
\ncline{P46A}{P46B}
\mput*{$F_{10}$}
\cnode(8,-20){3pt}{P47A}
\cnode(4,0){3pt}{P47B}
\ncline{P47A}{P47B}
\mput*{$G_{12}$}
\cnode(8,-20){3pt}{P48A}
\cnode(10,-4){3pt}{P48B}
\ncline{P48A}{P48B}
\mput*{$F_{15}$}
\cnode(8,-20){3pt}{P49A}
\cnode(14,-28){3pt}{P49B}
\ncline{P49A}{P49B}
\mput*{$F_{12}$}
\cnode(14,-20){3pt}{P50A}
\cnode(10,-4){3pt}{P50B}
\ncline{P50A}{P50B}
\mput*{$\!G_{14}$}
\cnode(14,-20){3pt}{P51A}
\cnode(18,-8){3pt}{P51B}
\ncline{P51A}{P51B}
\mput*{$F_{14}$}
\cnode(14,-20){3pt}{P52A}
\cnode(14,-28){3pt}{P52B}
\ncline{P52A}{P52B}
\mput*{$F_{15}$}
\cnode(20,-20){3pt}{P53A}
\cnode(18,-8){3pt}{P53B}
\ncline{P53A}{P53B}
\mput*{$F_{15}$}
\cnode(20,-20){3pt}{P54A}
\cnode(24,-2){3pt}{P54B}
\ncline{P54A}{P54B}
\mput*{$\!G_{13}$}
\cnode(20,-20){3pt}{P55A}
\cnode(14,-28){3pt}{P55B}
\ncline{P55A}{P55B}
\mput*{$F_{13}$}
\cnode(26,-22){3pt}{P56A}
\cnode(24,-2){3pt}{P56B}
\ncline{P56A}{P56B}
\mput*{$F_{15}$}
\cnode(26,-22){3pt}{P57A}
\cnode(28,2){3pt}{P57B}
\ncline{P57A}{P57B}
\mput*{$G_{11}$}
\cnode(26,-22){3pt}{P58A}
\cnode(14,-28){3pt}{P58B}
\ncline{P58A}{P58B}
\mput*{$F_{11}$}
\cnode(30,-26){3pt}{P59A}
\cnode(28,2){3pt}{P59B}
\ncline{P59A}{P59B}
\mput*{$F_{15}$}
\cnode(30,-26){3pt}{P60A}
\cnode(32,6){3pt}{P60B}
\ncline{P60A}{P60B}
\mput*{$G_{9}$}
\cnode(30,-26){3pt}{P61A}
\cnode(14,-28){3pt}{P61B}
\ncline{P61A}{P61B}
\mput*{$F_{9}$}
\cnode(34,-30){3pt}{P62A}
\cnode(32,6){3pt}{P62B}
\ncline{P62A}{P62B}
\mput*{$F_{15}$}
\cnode(34,-30){3pt}{P63A}
\cnode(36,10){3pt}{P63B}
\ncline{P63A}{P63B}
\mput*{$G_{7}$}
\cnode(34,-30){3pt}{P64A}
\cnode(14,-28){3pt}{P64B}
\ncline{P64A}{P64B}
\mput*{$F_{7}$}
\cnode(38,-34){3pt}{P65A}
\cnode(36,10){3pt}{P65B}
\ncline{P65A}{P65B}
\mput*{$F_{15}$}
\cnode(38,-34){3pt}{P66A}
\cnode(40,14){3pt}{P66B}
\ncline{P66A}{P66B}
\mput*{$G_{5}$}
\cnode(38,-34){3pt}{P67A}
\cnode(14,-28){3pt}{P67B}
\ncline{P67A}{P67B}
\mput*{$F_{5}$}
\cnode(42,-38){3pt}{P68A}
\cnode(40,14){3pt}{P68B}
\ncline{P68A}{P68B}
\mput*{$F_{15}$}
\cnode(42,-38){3pt}{P69A}
\cnode(44,18){3pt}{P69B}
\ncline{P69A}{P69B}
\mput*{$G_{3}$}
\cnode(42,-38){3pt}{P70A}
\cnode(14,-28){3pt}{P70B}
\ncline{P70A}{P70B}
\mput*{$F_{3}$}
\cnode(46,-42){3pt}{P71A}
\cnode(44,18){3pt}{P71B}
\ncline{P71A}{P71B}
\mput*{$F_{15}$}
\cnode(46,-42){3pt}{P72A}
\cnode(14,-28){3pt}{P72B}
\ncline{P72A}{P72B}
\mput*{$F_{1}$}
\cnode(-16,20){3pt}{P73A}
\cnode(44,18){3pt}{P73B}
\ncline{P73A}{P73B}
\mput*{$F_{1}$}
\cnode(-12,16){3pt}{P74A}
\cnode(40,14){3pt}{P74B}
\ncline{P74A}{P74B}
\mput*{$F_{3}$}
\cnode(-12,16){3pt}{P75A}
\cnode(44,18){3pt}{P75B}
\ncline{P75A}{P75B}
\mput*{$F_{2}$}
\cnode(-8,12){3pt}{P76A}
\cnode(36,10){3pt}{P76B}
\ncline{P76A}{P76B}
\mput*{$F_{5}$}
\cnode(-8,12){3pt}{P77A}
\cnode(40,14){3pt}{P77B}
\ncline{P77A}{P77B}
\mput*{$F_{4}$}
\cnode(-4,8){3pt}{P78A}
\cnode(32,6){3pt}{P78B}
\ncline{P78A}{P78B}
\mput*{$F_{7}$}
\cnode(-4,8){3pt}{P79A}
\cnode(36,10){3pt}{P79B}
\ncline{P79A}{P79B}
\mput*{$F_{6}$}
\cnode(0,4){3pt}{P80A}
\cnode(28,2){3pt}{P80B}
\ncline{P80A}{P80B}
\mput*{$F_{9}$}
\cnode(0,4){3pt}{P81A}
\cnode(32,6){3pt}{P81B}
\ncline{P81A}{P81B}
\mput*{$F_{8}$}
\cnode(4,0){3pt}{P82A}
\cnode(24,-2){3pt}{P82B}
\ncline{P82A}{P82B}
\mput*{$F_{11}$}
\cnode(4,0){3pt}{P83A}
\cnode(28,2){3pt}{P83B}
\ncline{P83A}{P83B}
\mput*{$F_{10}$}
\cnode(10,-4){3pt}{P84A}
\cnode(18,-8){3pt}{P84B}
\ncline{P84A}{P84B}
\mput*{$F_{13}$}
\cnode(10,-4){3pt}{P85A}
\cnode(24,-2){3pt}{P85B}
\ncline{P85A}{P85B}
\mput*{$F_{12}$}
\cnode(-18,-42){3pt}{P87A}
\cnode(46,-42){3pt}{P87B}
\ncline{P87A}{P87B}
\mput*{$G_{1}$}
\cnodeput*(-18.0,-42.0){P0}{${{30}}$}
\cnodeput(-18.0,-42.0){P0}{${{30}}$}
\cnodeput*(-14.0,-38.0){P1}{${{26}}$}
\cnodeput(-14.0,-38.0){P1}{${{26}}$}
\cnodeput*(-10.0,-34.0){P2}{${{22}}$}
\cnodeput(-10.0,-34.0){P2}{${{22}}$}
\cnodeput*(-6.0,-30.0){P3}{${{18}}$}
\cnodeput(-6.0,-30.0){P3}{${{18}}$}
\cnodeput*(-2.0,-26.0){P4}{${{14}}$}
\cnodeput(-2.0,-26.0){P4}{${{14}}$}
\cnodeput*(2.0,-22.0){P5}{${{10}}$}
\cnodeput(2.0,-22.0){P5}{${{10}}$}
\cnodeput*(8.0,-20.0){P6}{${{6}}$}
\cnodeput(8.0,-20.0){P6}{${{6}}$}
\cnodeput*(14.0,-20.0){P7}{${{2}}$}
\cnodeput(14.0,-20.0){P7}{${{2}}$}
\cnodeput*(20.0,-20.0){P8}{${{4}}$}
\cnodeput(20.0,-20.0){P8}{${{4}}$}
\cnodeput*(26.0,-22.0){P9}{${{8}}$}
\cnodeput(26.0,-22.0){P9}{${{8}}$}
\cnodeput*(30.0,-26.0){P10}{${{12}}$}
\cnodeput(30.0,-26.0){P10}{${{12}}$}
\cnodeput*(34.0,-30.0){P11}{${{16}}$}
\cnodeput(34.0,-30.0){P11}{${{16}}$}
\cnodeput*(38.0,-34.0){P12}{${{20}}$}
\cnodeput(38.0,-34.0){P12}{${{20}}$}
\cnodeput*(42.0,-38.0){P13}{${{24}}$}
\cnodeput(42.0,-38.0){P13}{${{24}}$}
\cnodeput*(46.0,-42.0){P14}{${{28}}$}
\cnodeput(46.0,-42.0){P14}{${{28}}$}
\cnodeput*(-16.0,20.0){P15}{${{29}}$}
\cnodeput(-16.0,20.0){P15}{${{29}}$}
\cnodeput*(-12.0,16.0){P16}{${{25}}$}
\cnodeput(-12.0,16.0){P16}{${{25}}$}
\cnodeput*(-8.0,12.0){P17}{${{21}}$}
\cnodeput(-8.0,12.0){P17}{${{21}}$}
\cnodeput*(-4.0,8.0){P18}{${{17}}$}
\cnodeput(-4.0,8.0){P18}{${{17}}$}
\cnodeput*(0.0,4.0){P19}{${{13}}$}
\cnodeput(0.0,4.0){P19}{${{13}}$}
\cnodeput*(4.0,0.0){P20}{${{9}}$}
\cnodeput(4.0,0.0){P20}{${{9}}$}
\cnodeput*(10.0,-4.0){P21}{${{5}}$}
\cnodeput(10.0,-4.0){P21}{${{5}}$}
\cnodeput*(18.0,-8.0){P22}{${{3}}$}
\cnodeput(18.0,-8.0){P22}{${{3}}$}
\cnodeput*(24.0,-2.0){P23}{${{7}}$}
\cnodeput(24.0,-2.0){P23}{${{7}}$}
\cnodeput*(28.0,2.0){P24}{${{11}}$}
\cnodeput(28.0,2.0){P24}{${{11}}$}
\cnodeput*(32.0,6.0){P25}{${{15}}$}
\cnodeput(32.0,6.0){P25}{${{15}}$}
\cnodeput*(36.0,10.0){P26}{${{19}}$}
\cnodeput(36.0,10.0){P26}{${{19}}$}
\cnodeput*(40.0,14.0){P27}{${{23}}$}
\cnodeput(40.0,14.0){P27}{${{23}}$}
\cnodeput*(44.0,18.0){P28}{${{27}}$}
\cnodeput(44.0,18.0){P28}{${{27}}$}
\cnodeput*(14.0,-28.0){P29}{${{1}}$}
\cnodeput(14.0,-28.0){P29}{${{1}}$}
\end{pspicture}
\end{center}
\caption{Our new construction of a complex extreme point $x^*$ for the subtour TSP polytope \LPNKb{2}, illustrated for $t=15$. Scaled edge values are shown: the label $F_i$ on an edge $e$ indicates that $x^*_e = F_i/F_t$. The symbol $G_i$ denotes $F_t - F_i$, i.e.\ an edge $e$ with $x^*_e = 1 - (F_i/F_t)$.
}\label{fig:constr}
\end{figure}

Let $F_i$ denote the $i$th Fibonacci number, where $F_1 = F_2 = 1$.
For a parameter $t \ge 3$, we denote the extreme point by $x^*$. The construction is given in the list below and pictured in \prettyref{fig:constr}.
\begin{itemize}
\setlength{\itemsep}{0pt}
\item For $i$ from 1 to $t$, an edge $(2i-1, 2i)$ of $x^*$-value 1
\item For $i$ from 2 to $t-1$, an edge $(1, 2i)$ of $x^*$-value $F_{t-i}/F_t$
\item An edge $(1, 2t)$ of $x^*$-value $1/F_t$
\item For $i$ from 3 to $t$, an edge $(2i-3, 2i-1)$ of $x^*$-value $F_{t-i+1}/F_t$
\item For $i$ from 3 to $t$, an edge $(2i-4, 2i-1)$ of $x^*$-value $1-F_{t-i+2}/F_t$
\item An edge $(2, 3)$ of $x^*$-value $F_{t-1}/F_t$
\item An edge $(2t-2, 2t)$ of $x^*$-value $1-1/F_t$
\end{itemize}

The support graph of $x^*$ has $2t$ vertices and $4t-3$ edges with fractionality $1/F_t$ and maximum degree $t$. Therefore, in order to prove \prettyref{theorem:family}, it suffices to show that $x^*$ is an extreme point solution.
\begin{prop}
The solution $x^*$ described above is an extreme point solution for \LPNKb{2}.
\end{prop}
\begin{proof}
With foresight, we write down the following family of $4t-3$ sets:
$$\mathcal{L} := \{\{i\}_{i=1}^{2t}, \{2i-1, 2i\}_{i=1}^t, \{1, \dotsc, 2i\}_{i=2}^{t-2}\}.$$

The plan of our proof is to first show that $x^*$ is the unique solution to $\{x(\delta(T))=2 \mid T \in \mathcal{L}\}$. It is easy to verify that $x^*$ indeed satisfies all these conditions, so let us focus on the harder task of showing that $x^*$ is the \emph{only} solution. (Note, we are not assuming that $x^*$ is feasible, so possibly $x^*(\delta(S)) < 2$ for some other sets, but we will deal with this later.)

A set $S$ is \emph{tight} for a solution $x$ if $x(\delta(S))=2$. Consider any solution which is tight for all sets in $\mathcal{L}$. We first need a simple lemma. For disjoint sets $S, T$, let $\delta(S:T)$ denote the set of edges with one end in $S$ and the other in $T$. 
\begin{lemma}
For some solution $x$, if $S, T$ are disjoint tight sets and $S \cup T$ is also tight, then $x(\delta(S:T))=1$.
\end{lemma}
\begin{proof}
We have $\delta(S) = \delta(S:T) \uplus \delta(S:V \bs S \bs T)$ and $\delta(T) = \delta(S:T) \uplus \delta(T:V \bs S \bs T)$. Also, $\delta(S \cup T) = \delta(S:V \bs S \bs T) \uplus \delta(T:V \bs S \bs T)$. Thus $2 = x(\delta(S))+x(\delta(T)) - x(\delta(S \cup T)) = 2x(\delta(S:T)).$
\end{proof}
Consider a hypothetical solution $x$ with $x(\delta(S)) = 2, \forall x \in \mathcal{L}$.
The lemma shows all edges $\{2i-1, 2i\}_{i=1}^t$ have $x$-value 1 (take $S = \{2i-1\}, T = \{2i\}$). Define $y_i$ equal to $x_{(2i+1, 2i+3)}$ for $i$ from 1 to $t-2$. The degree constraint at 3 (i.e., $x(\delta(3))=2$) forces $x_{(2, 3)} = 1-y_1$. The degree constraint at 2 forces $x_{(5, 2)} = y_1$. Note $\{1, \dotsc, 2t-2\}$ is tight since this set has the same constraint as $\{2t-1, 2t\}$. For $i$ from 1 to $t-2$, note that the sets $\delta(\{1, \dotsc, 2i\} : \{2i+1, 2i+2\})$ and $\delta(2i+1)$ differ only in that the former contains the edge $(2i+2, 1)$ and the latter contains the edges $\{(2i+1, 2i+2), (2i+1, 2i+3)\}$. Thus, using the lemma and degree constraint at $2i+1$, we see $x_{(2i+2, 1)}+x_{(2i+1,2i+3)}=y_i$. The degree constraint at $2i+2$ then forces $x_{(2i+2, 2i+5)}=1-y_i$ for $1 \le i \le t-3$. The degree constraint at $2t-2$ forces $x_{(1, 2t-2)}=1-y_{t-2};$ the degree constraint at $2t$ forces $x_{(1, 2t)}=y_{t-2}$. The degree constraint at $2t-1$ forces $y_{t-2} = y_{t-3}$, and the degree constraint at $2i+5$ forces $y_i = y_{i+1} + y_{i+2}$ for $i$ from 1 to $t-4$; together this shows $y_i = F_{t-1-i}\cdot y_{t-2}$ for $i$ from $t-4$ to 1 by induction. The degree constraint at 5 forces $2y_1 + y_2 = 1$, so $(2F_{t-2} + F_{t-3})y_{t-2}=1$ and consequently $y_{t-2} = 1/F_t$. Thus we conclude that $x=x^*$, as desired.

Now, we show $x^*$ is feasible using standard uncrossing arguments, plus the fact that $|\mathcal{L}| = 4t-3$. In \LPNKb{2}, the constraints for sets $S$ and $V \bs S$ are equivalent. Therefore, if we fix any root vertex $r \in V$, we may keep only the constraints for sets $S$ not containing $r$ without changing the LP. Correspondingly, we change $\mathcal{L}$ by complementing the sets that contain $r$, and it is easy to see $\mathcal{L}$ is a laminar family on $V \bs \{r\}$. (This is along the lines of the standard argument by Cornu\'{e}jols et al.\ \cite{CNF85}.) In fact $\mathcal{L}$ is a maximal laminar family, since any laminar family of nonempty subsets of $X$ contains at most $2|X|-1$ elements, for any set $X$.

Finally, suppose for the sake of contradiction that $x^*$ is not feasible, so there is a set $S$, with $r \not\in S$, having $x^*(\delta(S)) < 2$. Clearly $S \not\in \mathcal{L}$. Two sets $S, T$, neither containing $r$, \emph{cross} if all three of $S \bs T$, $T \bs S$, and $T \cap S$ are non-empty. Take $S$ with $x^*(\delta(S))<2$ such that $S$ crosses a minimal number of sets in $\mathcal{L}$. If $S$ crosses zero sets in $\mathcal{L}$, then $\mathcal{L} \cup \{S\}$ is laminar, but this is a contradiction since $S \not\in \mathcal{L}$ and, crucially, $\mathcal{L}$ was maximal. Otherwise, set $S$ crosses some tight set $T \in \mathcal{L}$, then since
$$2+2 > x^*(\delta(S)) + x^*(\delta(T)) \ge x^*(\delta(S \cup T)) + x^*(\delta(S \cap T)),$$
either $x^*(\delta(S \cup T)) < 2$ or $x^*(\delta(S \cap T)) < 2$. It is easy to verify that both $S \cup T$ and $S \cap T$ cross fewer sets of $\mathcal{L}$ than $S$, contradicting our choice of $S$.
\end{proof}

\subsection{Methodology}
\begin{figure}
\begin{center}
\begin{pspicture}(-5.5,-7)(10,10.5)
\psset{unit=1.8cm}
\cnode*(-1,5.5){3pt}{P39A}
\cnode*(-1,4.5){3pt}{P39B}
\ncline{P39A}{P39B}
\mput*{$1$}
\cnode*(-0.5,3.5){3pt}{P40A}
\cnode*(0.5,3){3pt}{P40B}
\ncline{P40A}{P40B}
\mput*{$1$}
\cnode*(-1.5,3.5){3pt}{P41A}
\cnode*(-2.5,3){3pt}{P41B}
\ncline{P41A}{P41B}
\mput*{$1$}
\cnode*(-1,4.5){3pt}{P42A}
\cnode*(-1.5,3.5){3pt}{P42B}
\ncline{P42A}{P42B}
\mput*{$\frac{1}{2}$}
\cnode*(-1.5,3.5){3pt}{P43A}
\cnode*(-0.5,3.5){3pt}{P43B}
\ncline{P43A}{P43B}
\mput*{$\frac{1}{2}$}
\cnode*(-0.5,3.5){3pt}{P44A}
\cnode*(-1,4.5){3pt}{P44B}
\ncline{P44A}{P44B}
\mput*{$\frac{1}{2}$}
\cnode*(-1,5.5){3pt}{P45A}
\cnode*(-2.5,3){3pt}{P45B}
\ncline{P45A}{P45B}
\mput*{$\frac{1}{2}$}
\cnode*(-2.5,3){3pt}{P46A}
\cnode*(0.5,3){3pt}{P46B}
\ncline{P46A}{P46B}
\mput*{$\frac{1}{2}$}
\cnode*(0.5,3){3pt}{P47A}
\cnode*(-1,5.5){3pt}{P47B}
\ncline{P47A}{P47B}
\mput*{$\frac{1}{2}$}
\cnode*(3,5){3pt}{P48A}
\cnode*(4,5){3pt}{P48B}
\ncline{P48A}{P48B}
\mput*{$1$}
\cnode*(4,5){3pt}{P49A}
\cnode*(5,5.5){3pt}{P49B}
\ncline{P49A}{P49B}
\mput*{$\frac{1}{2}$}
\cnode*(5,5.5){3pt}{P50A}
\cnode*(4,3){3pt}{P50B}
\ncline{P50A}{P50B}
\mput*{$1$}
\cnode*(4,3){3pt}{P51A}
\cnode*(3,3){3pt}{P51B}
\ncline{P51A}{P51B}
\mput*{$\frac{1}{2}$}
\cnode*(3,3){3pt}{P52A}
\cnode*(2,5.5){3pt}{P52B}
\ncline{P52A}{P52B}
\mput*{$1$}
\cnode*(2,5.5){3pt}{P53A}
\cnode*(3,5){3pt}{P53B}
\ncline{P53A}{P53B}
\mput*{$\frac{1}{2}$}
\cnode*(3,5){3pt}{P54A}
\cnode*(3.5,4){3pt}{P54B}
\ncline{P54A}{P54B}
\mput*{$\frac{1}{2}$}
\cnode*(4,5){3pt}{P56A}
\cnode*(3.5,4){3pt}{P56B}
\ncline{P56A}{P56B}
\mput*{$\frac{1}{2}$}
\cnode*(3.5,4){3pt}{P57A}
\cnode*(4,3){3pt}{P57B}
\ncline{P57A}{P57B}
\mput*{$\frac{1}{2}$}
\cnode*(3.5,4){3pt}{P58A}
\cnode*(3,3){3pt}{P58B}
\ncline{P58A}{P58B}
\mput*{$\frac{1}{2}$}
\cnode*(-2,2){3pt}{P59A}
\cnode*(0,2){3pt}{P59B}
\ncline{P59A}{P59B}
\mput*{$1$}
\cnode*(0,2){3pt}{P60A}
\cnode*(0,0){3pt}{P60B}
\ncline{P60A}{P60B}
\mput*{$\frac{1}{3}$}
\cnode*(0,0){3pt}{P61A}
\cnode*(-2,0){3pt}{P61B}
\ncline{P61A}{P61B}
\mput*{$1$}
\cnode*(-2,0){3pt}{P62A}
\cnode*(-2,2){3pt}{P62B}
\ncline{P62A}{P62B}
\mput*{$\frac{2}{3}$}
\cnode*(-2,2){3pt}{P63A}
\cnode*(-1.5,0.5){3pt}{P63B}
\ncline{P63A}{P63B}
\mput*{$\frac{1}{3}$}
\cnode*(-1.5,0.5){3pt}{P64A}
\cnode*(-1.5,1.5){3pt}{P64B}
\ncline{P64A}{P64B}
\mput*{$1$}
\cnode*(-1.5,1.5){3pt}{P65A}
\cnode*(-0.5,1.5){3pt}{P65B}
\ncline{P65A}{P65B}
\mput*{$\frac{2}{3}$}
\cnode*(-0.5,1.5){3pt}{P66A}
\cnode*(-0.5,0.5){3pt}{P66B}
\ncline{P66A}{P66B}
\mput*{$1$}
\cnode*(-0.5,0.5){3pt}{P67A}
\cnode*(-1.5,0.5){3pt}{P67B}
\ncline{P67A}{P67B}
\mput*{$\frac{1}{3}$}
\cnode*(-1.5,0.5){3pt}{P68A}
\cnode*(-2,0){3pt}{P68B}
\ncline{P68A}{P68B}
\mput*{$\frac{1}{3}$}
\cnode*(-1.5,1.5){3pt}{P69A}
\cnode*(0,2){3pt}{P69B}
\ncline{P69A}{P69B}
\mput*{$\frac{1}{3}$}
\cnode*(0,2){3pt}{P70A}
\cnode*(-0.5,1.5){3pt}{P70B}
\ncline{P70A}{P70B}
\mput*{$\frac{1}{3}$}
\cnode*(-0.5,0.5){3pt}{P71A}
\cnode*(0,0){3pt}{P71B}
\ncline{P71A}{P71B}
\mput*{$\frac{2}{3}$}
\cnode*(2,1){3pt}{P72A}
\cnode*(4.5,2.5){3pt}{P72B}
\ncline{P72A}{P72B}
\mput*{$1$}
\cnode*(4.5,2.5){3pt}{P73A}
\cnode*(5,1){3pt}{P73B}
\ncline{P73A}{P73B}
\mput*{$\frac{1}{3}$}
\cnode*(5,1){3pt}{P74A}
\cnode*(4.5,-0.5){3pt}{P74B}
\ncline{P74A}{P74B}
\mput*{$\frac{1}{3}$}
\cnode*(5,1){3pt}{P75A}
\cnode*(4.5,1.5){3pt}{P75B}
\ncline{P75A}{P75B}
\mput*{$\frac{1}{3}$}
\cnode*(4.5,2.5){3pt}{P76A}
\cnode*(3.5,1.5){3pt}{P76B}
\ncline{P76A}{P76B}
\mput*{$\frac{2}{3}$}
\cnode*(3.5,1.5){3pt}{P77A}
\cnode*(4.5,1.5){3pt}{P77B}
\ncline{P77A}{P77B}
\mput*{$1$}
\cnode*(4.5,1.5){3pt}{P78A}
\cnode*(3,1){3pt}{P78B}
\ncline{P78A}{P78B}
\mput*{$\frac{2}{3}$}
\cnode*(3,1){3pt}{P79A}
\cnode*(3.5,1.5){3pt}{P79B}
\ncline{P79A}{P79B}
\mput*{$\frac{1}{3}$}
\cnode*(3,1){3pt}{P80A}
\cnode*(3.5,0.5){3pt}{P80B}
\ncline{P80A}{P80B}
\mput*{$\frac{1}{3}$}
\cnode*(3.5,0.5){3pt}{P81A}
\cnode*(4.5,0.5){3pt}{P81B}
\ncline{P81A}{P81B}
\mput*{$\frac{2}{3}$}
\cnode*(4.5,0.5){3pt}{P82A}
\cnode*(3,1){3pt}{P82B}
\ncline{P82A}{P82B}
\mput*{$\frac{1}{3}$}
\cnode*(4.5,0.5){3pt}{P83A}
\cnode*(5,1){3pt}{P83B}
\ncline{P83A}{P83B}
\mput*{$1$}
\cnode*(4.5,-0.5){3pt}{P84A}
\cnode*(3.5,0.5){3pt}{P84B}
\ncline{P84A}{P84B}
\mput*{$1$}
\cnode*(4.5,-0.5){3pt}{P85A}
\cnode*(2,1){3pt}{P85B}
\ncline{P85A}{P85B}
\mput*{$\frac{2}{3}$}
\cnode*(2,1){3pt}{P86A}
\cnode*(3,1){3pt}{P86B}
\ncline{P86A}{P86B}
\mput*{$\frac{1}{3}$}
\cnode*(-0.5,-0.5){3pt}{P87A}
\cnode*(-2.5,-0.5){3pt}{P87B}
\ncline{P87A}{P87B}
\mput*{$1$}
\cnode*(-2.5,-0.5){3pt}{P88A}
\cnode*(-2.5,-3.5){3pt}{P88B}
\ncline{P88A}{P88B}
\mput*{$\frac{3}{4}$}
\cnode*(-2.5,-3.5){3pt}{P89A}
\cnode*(0.5,-3.5){3pt}{P89B}
\ncline{P89A}{P89B}
\mput*{$\frac{1}{2}$}
\cnode*(0.5,-3.5){3pt}{P90A}
\cnode*(0.5,-1.5){3pt}{P90B}
\ncline{P90A}{P90B}
\mput*{$1$}
\cnode*(0.5,-1.5){3pt}{P91A}
\cnode*(-0.5,-0.5){3pt}{P91B}
\ncline{P91A}{P91B}
\mput*{$\frac{1}{4}$}
\cnode*(-0.5,-0.5){3pt}{P92A}
\cnode*(-0.5,-1.5){3pt}{P92B}
\ncline{P92A}{P92B}
\mput*{$\frac{3}{4}$}
\cnode*(-0.5,-1.5){3pt}{P93A}
\cnode*(0.5,-1.5){3pt}{P93B}
\ncline{P93A}{P93B}
\mput*{$\frac{1}{4}$}
\cnode*(0.5,-3.5){3pt}{P94A}
\cnode*(-1,-3){3pt}{P94B}
\ncline{P94A}{P94B}
\mput*{$\frac{1}{2}$}
\cnode*(-1,-3){3pt}{P95A}
\cnode*(-2.5,-3.5){3pt}{P95B}
\ncline{P95A}{P95B}
\mput*{$\frac{1}{2}$}
\cnode*(-2.5,-3.5){3pt}{P96A}
\cnode*(-2,-2){3pt}{P96B}
\ncline{P96A}{P96B}
\mput*{$\frac{1}{4}$}
\cnode*(-2,-2){3pt}{P97A}
\cnode*(-2.5,-0.5){3pt}{P97B}
\ncline{P97A}{P97B}
\mput*{$\frac{1}{4}$}
\cnode*(-0.5,-1.5){3pt}{P98A}
\cnode*(-2,-2){3pt}{P98B}
\ncline{P98A}{P98B}
\mput*{$1$}
\cnode*(-2,-2){3pt}{P99A}
\cnode*(-1,-2){3pt}{P99B}
\ncline{P99A}{P99B}
\mput*{$\frac{1}{2}$}
\cnode*(-1,-2){3pt}{P100A}
\cnode*(-1,-3){3pt}{P100B}
\ncline{P100A}{P100B}
\mput*{$1$}
\cnode*(-1,-2){3pt}{P101A}
\cnode*(0.5,-1.5){3pt}{P101B}
\ncline{P101A}{P101B}
\mput*{$\frac{1}{2}$}
\cnode*(3.5,-2.5){3pt}{P112A}
\cnode*(2.5,-2.5){3pt}{P112B}
\ncline{P112A}{P112B}
\mput*{$\frac{1}{5}$}
\cnode*(3.5,-2.5){3pt}{P113A}
\cnode*(1.5,-3.5){3pt}{P113B}
\ncline{P113A}{P113B}
\mput*{$\frac{1}{5}$}
\cnode*(3.5,-2.5){3pt}{P114A}
\cnode*(3.5,-2){3pt}{P114B}
\ncline{P114A}{P114B}
\mput*{$1$}
\cnode*(3.5,-2.5){3pt}{P115A}
\cnode*(4.5,-2.5){3pt}{P115B}
\ncline{P115A}{P115B}
\mput*{$\frac{2}{5}$}
\cnode*(3.5,-2.5){3pt}{P116A}
\cnode*(5.5,-3.5){3pt}{P116B}
\ncline{P116A}{P116B}
\mput*{$\frac{1}{5}$}
\cnode*(1.5,-3.5){3pt}{P117A}
\cnode*(2,-1){3pt}{P117B}
\ncline{P117A}{P117B}
\mput*{$1$}
\cnode*(2,-1){3pt}{P118A}
\cnode*(2.5,-2.5){3pt}{P118B}
\ncline{P118A}{P118B}
\mput*{$\frac{4}{5}$}
\cnode*(2.5,-2.5){3pt}{P119A}
\cnode*(3,-1.5){3pt}{P119B}
\ncline{P119A}{P119B}
\mput*{$1$}
\cnode*(3,-1.5){3pt}{P120A}
\cnode*(3.5,-2){3pt}{P120B}
\ncline{P120A}{P120B}
\mput*{$\frac{2}{5}$}
\cnode*(3.5,-2){3pt}{P121A}
\cnode*(4,-1.5){3pt}{P121B}
\ncline{P121A}{P121B}
\mput*{$\frac{3}{5}$}
\cnode*(4,-1.5){3pt}{P122A}
\cnode*(4.5,-2.5){3pt}{P122B}
\ncline{P122A}{P122B}
\mput*{$1$}
\cnode*(4.5,-2.5){3pt}{P123A}
\cnode*(5,-1){3pt}{P123B}
\ncline{P123A}{P123B}
\mput*{$\frac{3}{5}$}
\cnode*(5,-1){3pt}{P124A}
\cnode*(5.5,-3.5){3pt}{P124B}
\ncline{P124A}{P124B}
\mput*{$1$}
\cnode*(5.5,-3.5){3pt}{P125A}
\cnode*(1.5,-3.5){3pt}{P125B}
\ncline{P125A}{P125B}
\mput*{$\frac{4}{5}$}
\cnode*(2,-1){3pt}{P126A}
\cnode*(5,-1){3pt}{P126B}
\ncline{P126A}{P126B}
\mput*{$\frac{1}{5}$}
\cnode*(5,-1){3pt}{P127A}
\cnode*(3,-1.5){3pt}{P127B}
\ncline{P127A}{P127B}
\mput*{$\frac{1}{5}$}
\cnode*(3,-1.5){3pt}{P128A}
\cnode*(4,-1.5){3pt}{P128B}
\ncline{P128A}{P128B}
\mput*{$\frac{2}{5}$}
\cnode*(2,5.5){3pt}{P129A}
\cnode*(5,5.5){3pt}{P129B}
\ncline{P129A}{P129B}
\mput*{$\frac{1}{2}$}
\rput(-3,4){(a)}
\rput(1.3,4){(b)}
\rput(-3,1){(c)}
\rput(1.3,1){(d)}
\rput(-3,-2){(e)}
\rput(1.3,-2){(f)}
\end{pspicture}
\end{center}
\caption{Six extreme points for the subtour TSP polytope \LPNKb{2} with extremal properties.}\label{fig:examples}
\end{figure}
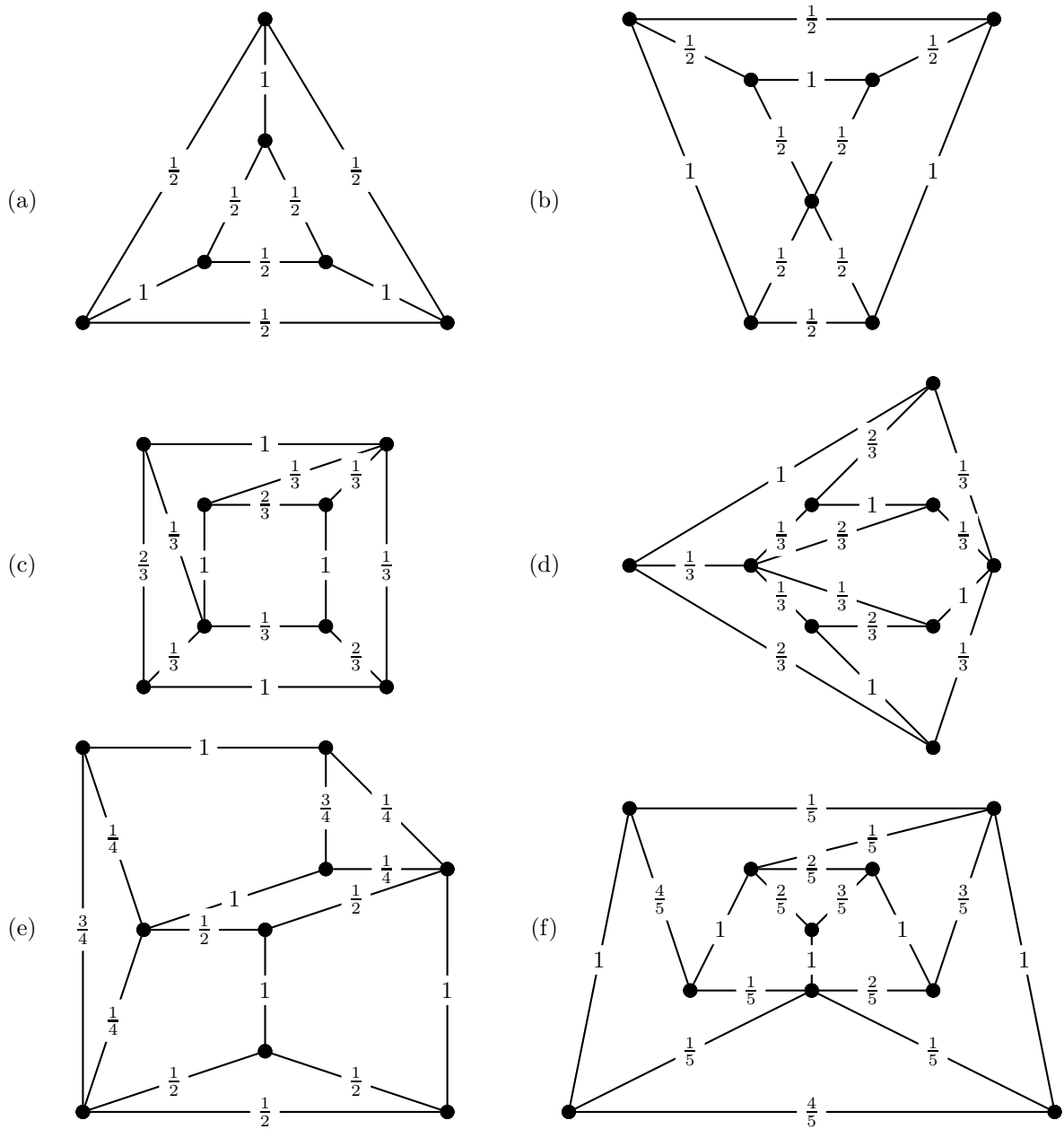

To investigate extreme points of \LPNKb{2}, we first used computational methods to try to find the most ``interesting" small examples. There are a number of properties that the support graph must have, e.g.\ no more than $2n-3$ edges, 3-vertex-connected (or else it is essentially a 2-sum of smaller solutions), and our method was to compute all extreme points on all such graphs. See Boyd~\cite{BB08,BE07} for more discussion of how these steps can be implemented. We used {\tt nauty}~\cite{McKay09} to generate the graphs, and the Maple package {\tt convex}~\cite{Franz09} to enumerate extreme points. The Maple package available at the time did not have a good interface for laying out graphs, so we created a procedure~\cite{Pweb09} to export the graphs to GeoGebra~\cite{Hohenwarter09}, which is well-suited for layout (and exporting for diagrams in this document). We found the following interesting examples, which are pictured in \prettyref{fig:examples}. Note ``unique" means unique up to graph isomorphism.
\begin{itemize}
\item[(a)] for $n \le 6$, there is a unique extreme point with denominator $\ge 2$
\item[(b)] for $n \le 7$, there is a unique extreme point with maximum degree $\ge 4$
\item[(c)] for $n \le 8$, there is a unique extreme point with denominator $\ge 3$
\item[(d)] for $n \le 9$, there is a unique extreme point with maximum degree $\ge 5$
\item[(e)] for $n \le 9$, there is a unique extreme point with denominator $\ge 4$
\item[(f)] for $n \le 10$, the maximum degree that occurs is 5 and the maximum denominator is 5; there is a unique solution on 10 vertices that attains both simultaneously
\end{itemize}
We found that there was some primal structure and dual structure to the 10-vertex example which was shared with the smaller examples (a) and (c); these observations led to the family described in \prettyref{sec:hk-cons}. We remark that the extreme points pictured, and more generally our new construction, do not coincide with the families of Boyd and Pulleyblank~\cite{BP91} or Cheung~\cite{Cheung03} for any choice of parameters.

\subsection{Discussion}
The construction given shows that extreme points on $n$ vertices of the Held-Karp relaxation may have maximum support degree as big as $n/2$ and fractionality as small as $1/F_{n/2}$, for even $n$. A natural question is whether these bounds are maximal. Boyd, with Benoit~\cite{BB08} and Elliott-Magwood~\cite{BE07}, has computed and posted online~\cite{B09} a list of all vertices of the subtour elimination polytope for up to 12 vertices. Filtering through that data, we find the following facts.

\begin{remark}For 11-vertex solutions, the largest maximum degree is 6, the largest denominator is 8, and of 11-vertex solutions with maximum degree 6, the maximum denominator is 5 which is uniquely attained. For 12-vertex solutions, the largest maximum degree is 6, the largest denominator is 9, and of 12-vertex solutions with maximum degree 6, the maximum denominator is 8 which is uniquely attained.
\end{remark}
Hence for even $n$, $F_{n/2}$ is not the maximum possible denominator. Based on the available data, we conjecture the following.
\begin{conj}
The maximum degree of extreme points on $n$ vertices is exactly $\lceil n/2 \rceil$.
\end{conj}
The best upper bound we are aware of is $n-3$, which follows from the fact that each basic solution has at most $2n-3$ edges, plus an easy argument to eliminate degree-2 vertices.

\subsection{Relation to Asymmetric TSP} Asymmetric TSP is the analogue of TSP for directed graphs: we are given a metric directed cost function on the complete digraph $(V, A)$, and seek a min-cost directed Hamiltonian cycle. Recently Asadpour et al.\ \cite{AG+10} obtained a breakthrough $O(\log n/\log \log n)$ approximation for this problem; its analysis uses the fact that extreme points of the natural LP relaxation
\begin{equation}\{ y \in \R_+^A : \forall \varnothing \neq U \subsetneq V, y(\delta^{\mathrm{out}}(U)) \ge 1 \}\label{eq:atsp}\tag{$\mathcal{A}$}\end{equation}
have denominator bounded by $2^{O(n \ln n)}$. Our undirected construction implies that for this directed variant, the extreme points attain denominator at least $2^{\Omega(n)}$.
\begin{prop}\label{prop:atsp}
For even $n \ge 6$ there are extreme points for \eqref{eq:atsp} on $n$ vertices with fractionality $1/F_{n/2}$ or smaller (and hence denominator at least $F_{n/2}$).
\end{prop}
\begin{proof}
The key is to note that \LPNKu{2} equals the projection of \eqref{eq:atsp} to $\R_+^E$ obtained by setting $x_{\{u, v\}} = y_{(u, v)}+y_{(v, u)}$ for all $\{u, v\} \in \tbinom{V}{2}$ (call this map \emph{dropping directions}). One direction is evident: given $y$, it has value at least 1 both coming into and coming out of every nontrivial cut set $U$, hence its undirected image $x$ has value at least 2 spanning the cut it defines, i.e.\ $x(\delta(U)) \ge 2$. Conversely, to show that for every $x \in \LPNKu{2}$, there is a $y \in$ \eqref{eq:atsp} of this type, just assign $y_{(u, v)}=y_{(v, u)}=x_{\{u, v\}}/2$ for all $\{u, v\} \in \tbinom{V}{2}$.

Now we prove \prettyref{prop:atsp}. Consider $x^*$ given by the construction, and consider the set of all $y$ in \eqref{eq:atsp} such that $y$ becomes $x^*$ when dropping directions. The argument in the previous paragraph establishes that this set is nonempty, and it is not hard to see this set is a face of \eqref{eq:atsp} since $x^*$ is an extreme point of $\LPNKu{2}$. Finally, let $y^*$ be any extreme point of this face. Our construction includes an edge $e$ with $x^*_e = 1/F_{n/2}$, hence at least one of the two arcs corresponding to $e$ has $y^*$-value in $(0, 1/F_{n/2}]$, giving the claimed result.

As a remark, the above proof leaves open the possibility that the extreme points $y^*$ for \eqref{eq:atsp} could have \emph{strictly worse} fractionality than $1/F_{n/2}$, but according to our computational experiments for $n=6, 8$, the worst-case fractionality for such $y^*$ is exactly $1/F_{n/2}$.
\end{proof}

\subsubsection{Integrality Gap} Several papers of Boyd and coauthors investigate TSP LP extreme points with the goal of lower-bounding the integrality gap, therefore it is natural to ask what integrality gap is implied by the construction given in this paper. It does not appear that our construction gives a good integrality gap lower bound; for $6, 8, 10, 12$ vertices we have computed that the integrality gap obtained is only $\frac{9}{8}, \frac{23}{21}, \frac{22}{20}, \frac{35}{32}$. (Specifically, this value is the least $t \ge 0$ such the extreme point is dominated by $t$ times a convex combination of indicator vectors of Hamiltonian cycles.)

\subsection*{Acknowledgement}
We thank Ashkan Aazami, Deeparnab Chakrabarty, Michel Goemans, Nick Harvey, and Jochen K\"onemann for useful discussions on these topics.

\bibliography{../huge}
\bibliographystyle{abbrv}

\appendix
\section{Hardness of \textsc{Path-Cover-of-Tree}}\label{app:pcot}
Our arguments are based on those of \cite{FJ81}, and also inspired by \cite{GVY93}, who used the same approach to prove \APX-hardness of a related packing problem. We reduce from minimum set cover in 3-uniform, 2-regular hypergraphs --- i.e.\ set cover with sets of size 3, each set appearing in exactly 2 sets --- which is equivalent to vertex cover in cubic graphs. The best known inapproximability ratio for this problem is about $\frac{100}{99}$, due to Chleb\'{\i}k and Chleb\'{\i}kov{\'a}~\cite{CC06}.

Here is the reduction. Let the instance of 3-uniform, 2-regular set cover be $(J, \mathcal{K})$ where $J$ is the ground set and $\mathcal{K}$ is the family of triples from $J$. Let $k = |\mathcal{K}|$ (so $|J| = 3k/2$) and denote the sets by $K_i = \{a[i], b[i], c[i]\}$ for $1 \le i \le k$ (so $a[i], b[i], c[i]$ are elements of $J$). The tree $T$ we construct for the \textsc{Path-Cover-of-Tree} instance has a root vertex $r$, a vertex $v_j$ for each $j \in J$, and two vertices $p_i, q_i$ for $1 \le i \le k$; $T$ has an edge $\{r, v_j\}$ for every $j \in J$, and the two edges $\{v_{a[i]}, p_i\}, \{v_{a[i]}, q_i\}$ for $1 \le i \le k$. Finally, we define the set $X$ to have the following $3k$ pairs: $\{p_i, q_i\}, \{p_i, v_{b[i]}\}, \{q_i, v_{c[i]}\}$ for $1 \le i \le k$.

\begin{claim}
$\OPT(T, X) = k + \OPT(J, \mathcal{K}).$
\end{claim}
(We speak of \textsc{Path-Cover-of-Tree} in terms of covering $E(T)$ instead of as a 2-connectivity problem.)
\begin{proof}
Let $\{K_i \mid i \in I\}$ be an optimal set cover, i.e.\ a $J$-covering subfamily of $\mathcal{K}$ such that $|I| = \OPT(J, \mathcal{K})$. Define $Y \subset X$ as follows: if $i \in I$ we put $\{p_i, v_{b[i]}\}$ and $\{q_i, v_{c[i]}\}$ into $Y$, and if $i \not\in I$ we put $\{p_i, q_i\}$ into $Y$. In either case, the corresponding paths in $T$ cover the edges incident to $p_i$ and $q_i$; and it is not hard to see that since $I$ is a set cover, all edges incident to $r$ are also covered. This proves $\OPT(T, X) \le 2|I| + (k-|I|) = k + \OPT(J, \mathcal{K}).$

The reverse inequality is similar. The only step needing pause is to consider whether $(T, X)$ always has an optimal solution $Y$ of the form generated by the above mapping (since then it can be reversed). Indeed, if $Y$ contains one or fewer of the 3 pairs $\{p_i, q_i\}, \{p_i, v_{b[i]}\}, \{q_i, v_{c[i]}\}$ then it must contain $\{p_i, q_i\}$ to cover the edges incident to $p_i$ and $q_i$; and if $Y$ contains two or more of the pairs, we can adjust such pairs to $\{p_i, v_{b[i]}\}$ and $\{q_i, v_{c[i]}\}$ without increasing $|Y|$ and without causing an edge of $T$ to become uncovered.
\end{proof}

Here are the calculations that show the reduction works. We have $\OPT(J, \mathcal{K}) \ge k/2$ (since we need to cover $3k/2$ points by triples), and by the result of \cite{CC06}, no polynomial-time algorithm can determine $\OPT(J, \mathcal{K})$ within additive error $\frac{k}{2 \cdot 99}$ on all instances, unless $\PP = \NP$. Hence, no polynomial-time algorithm can determine $\OPT(T, X)$ within the same additive error. Finally, since $\OPT(T, X) \le 2k$, we get an inapproximability ratio of $1+\frac{k}{2 \cdot 99}/2k = 1 + \frac{1}{396}$ for \textsc{Path-Cover-of-Tree}. However, if we actually look at the gap instances of \cite{CC06}, the same calculations give a slightly stronger ratio of $1 + \frac{1}{292.4}$.
\end{document}